\documentclass[pdflatex,iicol,sn-nature]{sn-jnl}

\usepackage{graphicx}
\usepackage{microtype}
\usepackage{amsmath,amssymb,bm}
\usepackage{booktabs,multirow,tabularx,array}
\usepackage{xcolor}
\usepackage{placeins}
\usepackage{enumitem}
\usepackage{longtable}
\usepackage{url}
\usepackage{manyfoot}

\graphicspath{{figures/}{figures/extended_data/}}
\newcommand{\runinhead}[1]{\par\medskip\noindent{\bfseries\itshape #1.}\hspace{0.5em}\ignorespaces}

\begin{document}

\title[Full Bayesian inference of spinning EMRIs]{Machine learning enables full Bayesian inference for spinning extreme-mass-ratio inspirals}

\author[1,2,3]{\fnm{Bo} \sur{Liang}}
\equalcont{These authors contributed equally to this work.}
\author*[1,3]{\fnm{Minghui} \sur{Du}}
\equalcont{These authors contributed equally to this work.}
\author*[4]{\fnm{Zhenwei} \sur{Lyu}}
\equalcont{These authors contributed equally to this work.}
\author[3]{\fnm{Hanlin} \sur{Song}}
\author[5]{\fnm{Jiakai} \sur{Zhang}}
\author[5]{\fnm{Weichen} \sur{Qin}}
\author[6]{\fnm{Chia-Jui} \sur{Chou}}
\author[6]{\fnm{Yi} \sur{Yang}}
\author[5]{\fnm{Jingyi} \sur{Yu}}
\author[1,3,8]{\fnm{Ziren} \sur{Luo}}
\author*[2]{\fnm{Mingming} \sur{Sun}}\email{mmsun@bimsa.cn}
\author*[1,3,7,8]{\fnm{Peng} \sur{Xu}}\email{xupeng@imech.ac.cn}
\author*[1,7]{\fnm{Yue-Liang} \sur{Wu}}\email{ylwu@ucas.edu.cn}

\affil[1]{\orgdiv{Taiji Laboratory for Gravitational Wave Universe (Beijing/Hangzhou)}, \orgname{University of Chinese Academy of Sciences}, \orgaddress{\city{Beijing}, \postcode{100049}, \country{China}}}
\affil[2]{\orgdiv{AGI Lab}, \orgname{Beijing Institute of Mathematical Sciences and Applications}, \orgaddress{\city{Beijing}, \postcode{101408}, \country{China}}}
\affil[3]{\orgdiv{Center for Gravitational Wave Experiment, National Microgravity Laboratory}, \orgname{Institute of Mechanics, Chinese Academy of Sciences}, \orgaddress{\city{Beijing}, \postcode{100190}, \country{China}}}
\affil[4]{\orgdiv{Leicester International Institute}, \orgname{Dalian University of Technology}, \orgaddress{\city{Panjin}, \postcode{124221}, \country{China}}}
\affil[5]{\orgdiv{School of Information Science and Technology}, \orgname{ShanghaiTech University}, \orgaddress{\city{Shanghai}, \postcode{201210}, \country{China}}}
\affil[6]{\orgdiv{School of Physical Science and Technology}, \orgname{ShanghaiTech University}, \orgaddress{\city{Shanghai}, \postcode{201210}, \country{China}}}
\affil[7]{\orgdiv{International Centre for Theoretical Physics Asia-Pacific}, \orgname{University of Chinese Academy of Sciences}, \orgaddress{\city{Beijing}, \postcode{100049}, \country{China}}}
\affil[8]{\orgdiv{Key Laboratory of Gravitational Wave Precision Measurement of Zhejiang Province, Hangzhou Institute for Advanced Study}, \orgname{University of Chinese Academy of Sciences}, \orgaddress{\city{Hangzhou}, \postcode{310024}, \country{China}}}


\abstract{\unboldmath Extreme-mass-ratio inspirals (EMRIs), stellar-mass compact objects spiralling into massive black holes, promise to measure the masses and spins of the central black holes, reveal the strong-field environment around them and gauge cosmic distances via millihertz gravitational-wave observations. Yet most forecasts for the parameter inference of these systems assume known source parameters, and broad-prior analyses have addressed only simplified versions of the problem. The joint inversion of all 14 parameters of a spinning source from the complete prior has remained unsolved. Here we solve it for the augmented analytic kludge model: generative networks narrow the complete prior and reveal a ring of degenerate spin directions, and the exact detector likelihood determines the posterior. For a simulated one-year LISA--Taiji observation, within one ring sector we obtain, for the first time, a 14-parameter posterior of a spinning EMRI from the complete prior, the most precise yet obtained for such a source from broad priors. The redshifted mass and spin of the massive black hole are determined to two parts in $10^{5}$, the mass two orders of magnitude beyond any electromagnetic measurement. The compact object's redshifted mass is measured to one part in $10^{4}$, and its eccentric, retrograde orbit discriminates between capture channels. The 0.8\% distance and sub-square-degree localization are sufficient for dark-siren cosmology. Year-long coherent analysis resolves the inspiral driven by radiation reaction itself.}

\maketitle

\begin{figure*}[t]
  \centering
  \includegraphics[width=0.9\textwidth]{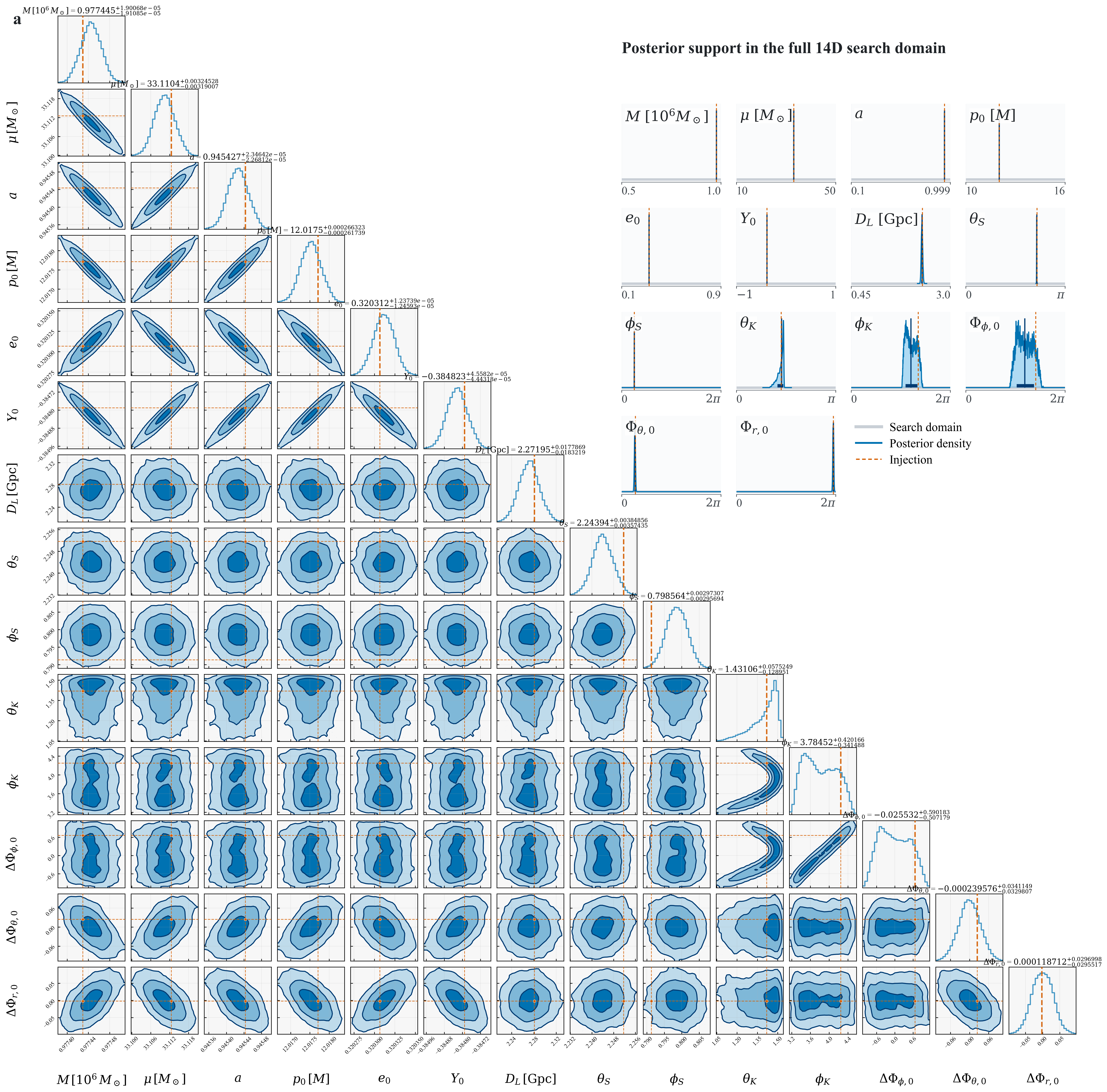}
    \caption{\textbf{Fourteen-parameter posterior of a spinning EMRI from one year of joint LISA--Taiji data.} Lower left, one- and two-dimensional marginal densities from parallel-tempered Markov chain Monte Carlo in which all fourteen parameters, including the luminosity distance, are sampled under the exact two-detector likelihood, starting from the completed thirteen-parameter analysis (Methods). Contours enclose 39\%, 86\% and 99\% of the two-dimensional probability, and labels give the median and central 68\% interval of each parameter. Orbital phases are shown as offsets from their circular means. Upper right, each one-dimensional marginal drawn within its complete prior range (grey band), showing the contraction from prior to posterior. Orange dashed lines mark the injected values, added after sampling and not used by the sampler.}
  \label{fig:posterior_geometry}
\end{figure*}
\begin{figure*}[t]
\centering
\includegraphics[width=\textwidth]{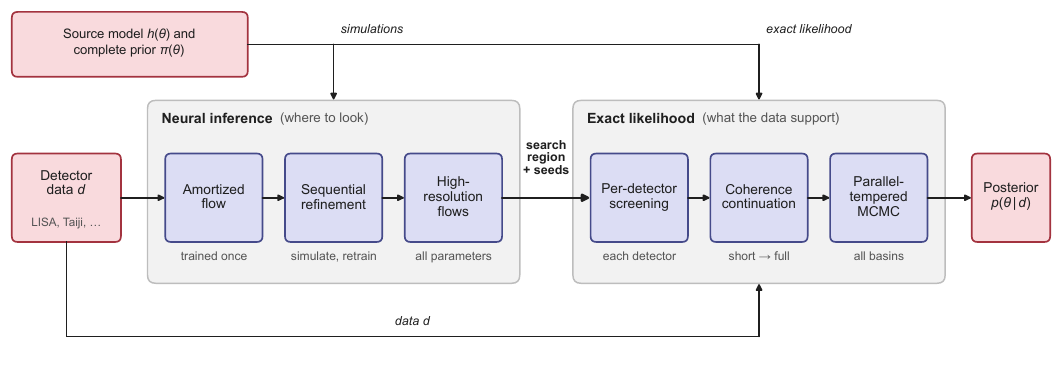}
\caption{\textbf{The NEBULA-EMRI framework.} Data from a network of detectors pass through a neural-inference stage, which decides where to look, and an exact-likelihood stage, which decides what the data support. The networks are trained on simulations from the source model and prior and pass only a search region and starting points (seeds) across the boundary between the stages. The exact stage evaluates the likelihood of the data directly. Candidates are screened independently in each detector, followed as the coherent integration time grows from a short segment to the full observation, and combined across basins by parallel-tempered Markov chain Monte Carlo, whose cold chain gives the posterior. No neural density enters the likelihood or weights the posterior.}
\label{fig:pipeline}
\end{figure*}
In general relativity, an isolated, stationary black hole is completely characterized by its mass, spin and electric charge~\cite{chrusciel2012uniqueness}. Because astrophysical black holes are expected to carry negligible charge, they are described by their mass and spin alone, through the rotating vacuum solution that Kerr found more than six decades ago~\cite{kerr1963rotating,carter1971nohair}. In galactic nuclei, these two quantities record how massive black holes grew~\cite{reynolds2021spin}, yet both remain difficult to measure. Horizon-scale imaging and stellar orbits determine the masses of only a few nearby objects, to between about ten per cent and a few tenths of a per cent~\cite{eht2019m87,gravity2022mass}, whereas spins inferred from electromagnetic observations depend on models of the accretion flow~\cite{reynolds2021spin}. Extreme-mass-ratio inspirals (EMRIs) offer an independent route to measuring both, in quiescent and active galaxies alike, and they also determine the source distance without relying on the cosmic distance ladder. A stellar-mass compact object spiralling into a massive black hole provides a natural test-particle experiment in strong gravity. Over tens of thousands of relativistic orbits, the phase of its gravitational radiation encodes the multipole moments of the massive black hole's spacetime, as first shown by Ryan~\cite{ryan1995multipoles}. Space-based detectors such as LISA~\cite{colpi2024lisa,esa2025lisa} and Taiji~\cite{hu2017taiji} can track this phase coherently for a year or longer, over which it accumulates some $10^{5}$ radians. Because the phase is the time integral of the orbital frequencies, a fractional frequency error of $10^{-5}$ builds up about one radian of dephasing over the observation. At the signal-to-noise ratios of 20 or more expected for detected EMRIs~\cite{babak2017science}, phase differences of order the inverse signal-to-noise ratio are resolvable, so this dephasing is readily distinguished and the orbital frequencies can be constrained to about one part in $10^{5}$. A single EMRI could therefore determine the mass and spin of a massive black hole to $10^{-4}$ or better~\cite{barack2004lisa,babak2017science}, reveal how the compact object was captured~\cite{amaro2018relativistic,pan2021agn}, probe dark matter and new fundamental fields around the black hole~\cite{eda2013dmspike,duque2024axion,maselli2022fields} and serve as a dark siren for cosmology~\cite{schutz1986hubble,cai2017gwphysics,laghi2021cosmology}. No other observable offers this reach: binary pulsars probe orbits moving at a few thousandths of the speed of light~\cite{kramer2006science}, and merging stellar-mass black holes spend only a few cycles in the strong field~\cite{abbott2016gw150914}.
These prospects make EMRIs a defining target of LISA and Taiji. According to their published plans and roadmaps, the science operations of the two missions are expected to overlap from around 2035~\cite{colpi2024lisa,esa2025lisa,hu2017taiji}, so every EMRI observed during this period will be recorded by two independent instruments with distinct orbits and responses, allowing each inferred feature to be cross-checked~\cite{ruan2020network}. Waveform modelling has advanced in parallel~\cite{wardell2023secondorder,chua2021relativistic,katz2021few}. The augmented analytic kludge (AAK), for example, maps its orbital frequencies onto those of Kerr geodesics~\cite{chua2017aak}. Progress on the inverse problem of inferring the source parameters from the data has lagged behind, even though all of these science goals depend on solving it.

For a generic inspiral into a spinning black hole, this inverse problem is exceptionally challenging. Fourteen parameters must be inferred jointly. The three fundamental frequencies of Kerr orbits, modulated by the detectors' motion, generate a dense spectral forest~\cite{drasco2006kerr}. Because fractional offsets of only $10^{-5}$ accumulate a radian of phase within a year, the likelihood is confined to thin, curved manifolds within a vast prior that is full of distant secondary maxima~\cite{chua2022nonlocal}, so a fully coherent template search would need $\sim10^{40}$ templates~\cite{gair2004eventrates}. Spin both complicates the problem and makes it worth solving. Frame dragging, measured only to about ten per cent in Earth orbit~\cite{ciufolini2004lense,everitt2011gpb}, here shapes every orbit and entangles the inclination, the spin orientation and the orbital phases. A missed likelihood basin or a spurious match to a subset of harmonics biases the inferred masses, spin and orbital parameters~\cite{chua2022nonlocal}, and can mask or mimic the small dephasings arising from the astrophysical environment of the black hole~\cite{cardoso2022environments,speri2023accretion}. Neural inference now analyses ground-based detections in real time~\cite{dax2021realtime,dingobns2025nature}, and conventional search algorithms and machine-learning methods have steadily widened the priors from which EMRIs can be recovered~\cite{strub2025search,cole2026sequential,liang2024rapid,liang2025fmmcmc,zou2026endtoend}. These analyses, however, either return only point estimates or obtain posteriors for simplified versions of the problem, neglecting the black-hole spin, inferring only a subset of the parameters from broad priors, or analysing at most half a year of data (Methods and Extended Data Table~\ref{tab:prior_work}). Real observations permit none of these simplifications: massive black holes are expected to spin~\cite{reynolds2021spin}, no source parameter is known in advance, and the full precision of an EMRI requires a year or more of coherent phase tracking. Yet a year-long Bayesian posterior over all 14 parameters of a generic spinning source, obtained from its complete prior, has remained out of reach.

Here we solve this inversion for a spinning EMRI described by the AAK waveform model, using Neural-to-Exact Bayesian Unified Likelihood Analysis for EMRIs (NEBULA-EMRI), which proceeds in two stages (Fig.~\ref{fig:pipeline}). In the neural-inference stage, an amortized network trained over the complete prior constrains the masses, spin magnitude, orbital elements and distance in under a second. Networks retrained on simulations in successively narrower regions, aided by an intermediate likelihood search, then extend the constraints to all 14 parameters and reveal a ring-shaped degeneracy in the spin direction. The exact-likelihood stage follows one sector of this ring. Candidates are first screened with 30 days of data from each detector separately, and only those supported by both LISA and Taiji are kept. The coherent integration time is then extended step by step from 30 days to a full year, and parallel-tempered Markov chain Monte Carlo samples all 14 parameters under the exact one-year two-detector likelihood. No neural density enters the likelihood or weights the posterior. The networks define only the search region and the starting points.

We apply NEBULA-EMRI to a simulated one-year LISA--Taiji observation of an EMRI whose parameters were drawn at random from the complete prior and which lies at a luminosity distance of 2.3~Gpc (redshift $z\approx0.4$; Extended Data Table~\ref{tab:final_intervals}). At this distance the signal reaches signal-to-noise ratios of 46 in LISA and 123 in Taiji, or 131 for the two-detector network. NEBULA-EMRI yields a year-long posterior over all 14 parameters (Fig.~\ref{fig:posterior_geometry}). To our knowledge, it is the first joint Bayesian posterior over all 14 parameters of a spinning EMRI recovered from its complete prior, and the most precise yet obtained for such a source from broad priors. This precision gives the inversion its scientific value. A single year of data determines the redshifted mass and spin of a massive black hole far more precisely than electromagnetic methods allow, constrains how the compact object was captured, and localizes the source well enough for dark-siren cosmology.

\begin{table*}[t]
\centering
\caption{\textbf{Physical reach of the year-long joint LISA--Taiji posterior.} Precisions are central 68\% fractional half-widths unless stated otherwise (Methods).}
\label{tab:physics_reach}
\small
\begin{tabularx}{\textwidth}{>{\raggedright\arraybackslash\hsize=0.85\hsize}X l >{\raggedright\arraybackslash\hsize=1.2\hsize}X >{\raggedright\arraybackslash\hsize=0.95\hsize}X}
\toprule
Quantity & Precision & Physics enabled & Reference point\\
\midrule
Massive-black-hole mass $M$ (detector frame) & $1.9\times10^{-5}$ & Demography and growth of massive black holes & Best electromagnetic measurement, $2.5\times10^{-3}$ (Sgr~A*)~\cite{gravity2022mass}\\
Spin $a$ & $2.4\times10^{-5}$ & Frame dragging and growth history (accretion versus mergers)~\cite{reynolds2021spin} & Spins from electromagnetic observations, which depend on accretion-flow models~\cite{reynolds2021spin}\\
Compact-object mass $\mu$ (detector frame) & $9.7\times10^{-5}$ ($0.003\,M_\odot$) & Stellar-mass black holes in galactic nuclei & Component masses of merging binaries, tens of per cent~\cite{gwtc3}\\
Orbital elements $(p_0,e_0,Y_0)$ & $(2.2,\,3.9,\,12)\times10^{-5}$ & Capture channel: two-body relaxation versus accretion-disk migration~\cite{amaro2018relativistic,pan2021agn} & Channels predict distinct eccentricity and inclination distributions~\cite{pan2021agn}\\
Luminosity distance $D_L$ & $8\times10^{-3}$ & Dark-siren measurement of the Hubble constant~\cite{laghi2021cosmology} & Per-cent-level $H_0$ forecast from loud EMRIs~\cite{laghi2021cosmology}\\
Sky position $(\theta_S,\phi_S)$ & $4\times10^{-3}$, $3\times10^{-3}$\,rad & Host-galaxy identification and electromagnetic follow-up & Region of a fraction of a square degree\\
Joint 14-parameter covariance & --- & Correlated uncertainties of all parameters, including the spread between basins & Unavailable from point estimates\\
\bottomrule
\end{tabularx}
\end{table*}

\runinhead{What a year of data measures}

With NEBULA-EMRI, the year-long posterior determines the detector-frame mass and the spin of the massive black hole to fractional precisions of 19 and 24 parts per million, respectively. It also constrains the detector-frame mass of the compact object and the three orbital elements (initial semi-latus rectum, eccentricity and cosine of the inclination) to fractional precisions between $2\times10^{-5}$ and $1.2\times10^{-4}$, the luminosity distance to 0.8\% and the sky position to a fraction of a square degree (68\% credibility; Fig.~\ref{fig:posterior_geometry} and Table~\ref{tab:physics_reach}). Forecasts that assume known source parameters have long predicted such precisions for EMRIs~\cite{barack2004lisa,babak2017science}. Here they are achieved by a global analysis that starts from the complete prior. Multiplied over all 14 parameters, the ratio of the prior range to the 68\% posterior width approaches $10^{40}$, comparable to the number of templates once estimated for a fully coherent search~\cite{gair2004eventrates}. No previous broad-prior analysis of a spinning EMRI has achieved comparable posterior precision. The only earlier broad-prior posterior for a spinning source, obtained in our previous study~\cite{liang2025fmmcmc}, inferred four intrinsic parameters from 60 days of single-detector data, whereas 14-parameter searches have so far returned only point estimates. The most recent of these, a coherent half-year single-detector search using the original analytic kludge waveform, reported mass and spin errors one to two orders of magnitude larger than those obtained here~\cite{zou2026endtoend} (Methods and Extended Data Table~\ref{tab:prior_work}). Ten of the fourteen injected values, including all seven that govern the orbital dynamics and signal amplitude, lie within their 68\% intervals, as expected for calibrated intervals (Methods).

\begin{figure*}[t]
  \centering
  \includegraphics[width=0.95\textwidth]{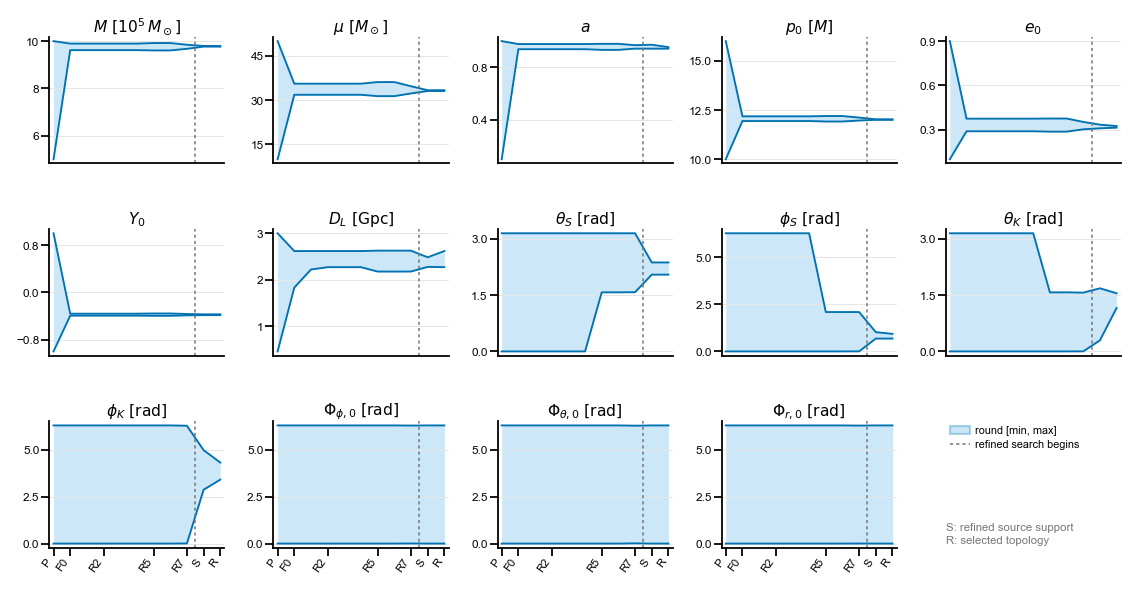}
  \caption{\textbf{Coordinate-wise support from the complete prior to the region passed to the exact-likelihood stage.} Shaded bands span the lower and upper bounds of the simulation support of each parameter, from the complete prior (P) through the amortized network (F0) and the sequential rounds R1--R7 to the high-resolution simulation domain (S) and the region handed to the exact-likelihood stage (R). The six intrinsic parameters and the luminosity distance contract at F0, whereas the sky position and spin orientation keep their complete ranges until R5. The step in these angles at R5 reflects the selection of one orientation cell (Methods). The orbital phases retain their complete periodic ranges throughout. The dotted line marks the start of high-resolution refinement. Angles are in radians.}
  \label{fig:support_evolution}
\end{figure*}

Each of these measurements has a distinct physical implication (Table~\ref{tab:physics_reach}). At a gigaparsec distance, the detector-frame mass of the black hole is determined two orders of magnitude more precisely than the best electromagnetic measurement of any massive black hole, that of the Galactic Centre~\cite{gravity2022mass}. Its spin, which shapes every orbit through frame dragging and records whether the black hole grew mainly by coherent accretion or through mergers~\cite{reynolds2021spin}, is measured to a fractional precision of $2.4\times10^{-5}$. The detector-frame mass of the compact object is measured to a fractional precision of about $10^{-4}$, three orders of magnitude more precisely than the component masses of merging binary black holes~\cite{gwtc3}. Combined with a redshift estimate, this enables direct measurement of the source-frame masses of stellar-mass black holes in galactic nuclei. The eccentricity and inclination of the compact object's orbit, measured to parts in $10^{5}$--$10^{4}$, discriminate between capture through two-body relaxation, which yields eccentric orbits of any inclination, and migration through an accretion disk, which yields nearly circular, prograde orbits aligned with the disk~\cite{amaro2018relativistic,pan2021agn}. The distance and sky position narrow the candidate host galaxies to a small volume, reaching the precision needed for electromagnetic follow-up and for dark-siren measurements of the Hubble constant independent of the cosmic distance ladder~\cite{schutz1986hubble,abbott2017hubble,laghi2021cosmology}. The data also constrain the angle between the spin axis and the line of sight. For EMRIs in active galaxies, comparing this angle with the inclinations of accretion disks and jets measured by telescopes~\cite{reynolds2021spin} could reveal whether black holes and their disks are aligned, a key question for how black holes grow and shape their host galaxies.

\begin{figure*}[t]
  \centering
  \includegraphics[width=0.9\textwidth]{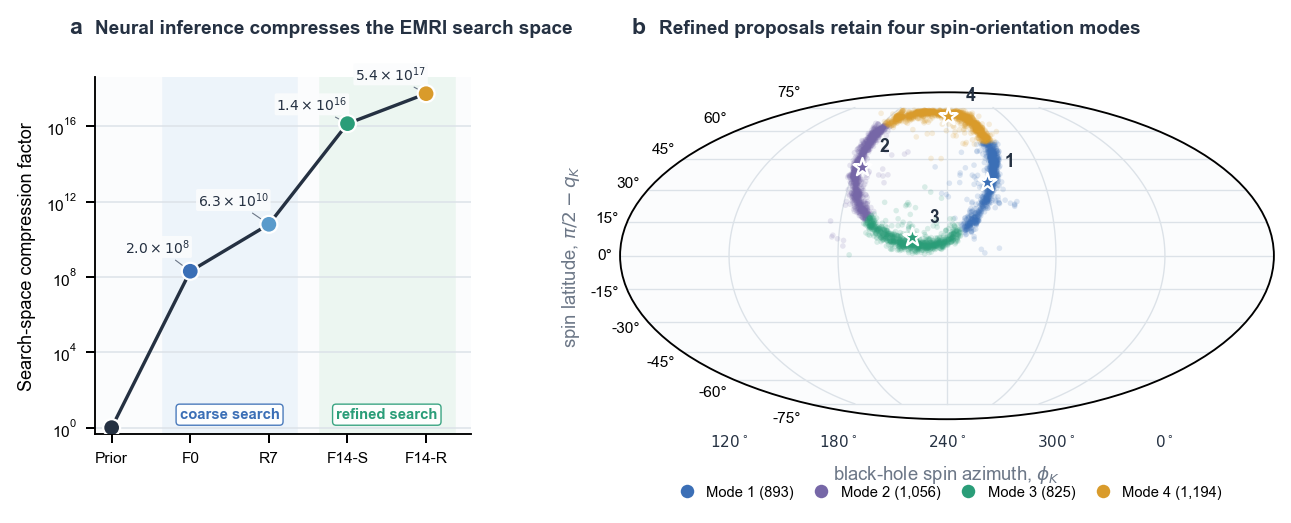}
  \caption{\textbf{The neural-inference stage shrinks the search region and reveals a ring of spin directions.} \textbf{a}, Contraction of the search region relative to the complete 14-parameter prior, that is, the prior volume divided by the volume of the region, after the amortized network (F0) and the last sequential round (R7), for the high-resolution simulation domain (F14-S) and for the region passed to the exact-likelihood stage (F14-R). \textbf{b}, Spin directions of the 3,968 high-resolution proposals within the spin bounds, in ecliptic coordinates (Mollweide projection). The proposals form a ring centred close to the line of sight, and $k$-means clustering divides it into four sectors with 893, 1,056, 825 and 1,194 proposals, and stars mark the sector centres. The exact-likelihood stage follows sector 3 (Methods).}
  \label{fig:neural_topology}
\end{figure*}

\runinhead{Narrowing the complete prior}

Obtaining such measurements from an unknown signal first requires locating the source across the complete prior, long regarded as the central computational obstacle of EMRI science~\cite{babak2010mldc}. Amortized inference turns this search into a single network evaluation: trained once on simulations spanning the complete prior, a network maps a year of LISA--Taiji data, in under a second, to a probability distribution over the masses, spin magnitude, orbital elements and distance, whose central intervals bound the region of parameter space consistent with the data. On 2,000 independent test signals with different source parameters and signal-to-noise ratios, the region contracts more strongly as the signal-to-noise ratio increases, as measured by the ratio of the complete ranges of the network's targets to the extent of the region (Methods), while its central-$3\sigma$ part still contains the true source in 98\% of cases (Extended Data Fig.~\ref{fig:amortized_snr}). For the source analysed here, the network identifies the nature of the system within a second: a rapidly spinning black hole of about a million solar masses and a compact object of 32--36 solar masses on an eccentric, retrograde orbit, at 1.8--2.6~Gpc (Fig.~\ref{fig:support_evolution}). Such an amortized network could scan the LISA and Taiji data streams, which will contain many thousands of overlapping signals~\cite{colpi2024lisa}, rank EMRI candidates by signal-to-noise ratio and pass them on for sequential refinement and exact-likelihood analysis, providing the front end of a pipeline for cataloguing massive black holes and the orbits of the compact objects they capture.
\begin{figure*}[t]
  \centering
  \includegraphics[width=0.9\textwidth]{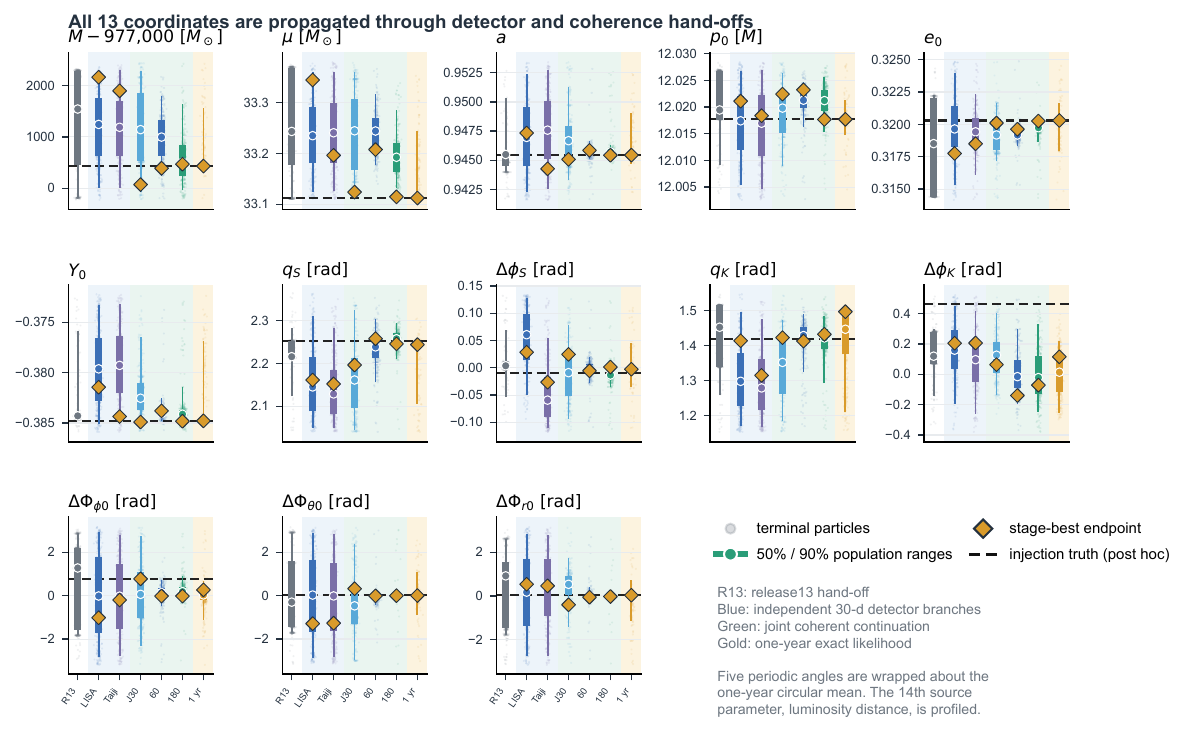}
  \caption{\textbf{Detector checks and longer coherence carry the candidate basins to one year.} Final populations of the exact-likelihood searches for each of the 13 sampled parameters, from the 13-parameter refinement (R13) through separate 30-day searches in LISA and Taiji (blue), joint searches over 30, 60 and 180 days (green) and the one-year exact likelihood. Thick and thin bars show the central 50\% and 90\% of each population, and diamonds mark the best point of each stage. Dashed lines show the injected values, added after the analysis. Five periodic angles are shown relative to their one-year circular means, and the luminosity distance is profiled. The 360-day stage is not shown, for clarity.}
  \label{fig:exact_handoffs}
\end{figure*}

Sequential refinement then extends the neural-inference stage to all 14 parameters, adding the sky position, the spin direction and the orbital phases. The network is retrained on new simulations within the current region over several rounds, followed by two high-resolution networks centred on the result of an intermediate likelihood search (Methods). Together they shrink the region to be searched by a factor of $5.4\times10^{17}$ relative to the complete prior (Figs.~\ref{fig:support_evolution} and~\ref{fig:neural_topology}a). This contraction factor is the volume of the complete prior divided by that of the search region. Because both are boxes in the 14 parameters, it equals the ratio of each parameter's prior range to its retained range, multiplied over all 14 parameters (Methods). The output is a map of all solutions consistent with the data rather than a single best estimate, and it reveals a clear pattern: the allowed spin directions form a ring centred close to the line of sight (Fig.~\ref{fig:neural_topology}b). The ring has a simple physical origin. If the whole system is rotated about the line of sight, only the polarization of the waves changes, and this change can be largely undone by shifting the orbital phases. The data therefore constrain the angle between the spin axis and the line of sight but leave the azimuth around it largely free. The network learned this symmetry from the simulations without being given it explicitly. Detectors with different orientations measure the polarization better and thus constrain the position along the ring, one reason why a network of detectors is valuable~\cite{ruan2020network}. Mapping such degeneracies before the final analysis matters, because a structure that goes unnoticed biases the spin direction and, through correlations, the masses, spin and orbital parameters~\cite{chua2022nonlocal}.

\runinhead{Coherence across detectors and a year of data}

The feature that makes EMRIs so precise also makes them hard to find. Over one year the phase of the signal accumulates about $10^{5}$ radians, so a template whose orbital frequencies deviate from the true values by a few parts in $10^{5}$ loses phase coherence with the data. A one-year search initialized directly from the region identified by the networks therefore becomes trapped in secondary likelihood maxima, none of which fits the data better than noise alone (Methods). The analysis instead starts from 30 days of data and asks for what any claimed signal needs: independent confirmation. Differential evolution, a likelihood-based optimizer, is run separately on the LISA and the Taiji data, and only candidates that fit both better than noise alone are kept (Fig.~\ref{fig:exact_handoffs}). Two detectors with different orbits, orientations and noise are unlikely to be misled in the same way by a false match to part of the signal.

The coherent integration time is then increased step by step from 30 days to a full year (Fig.~\ref{fig:exact_handoffs}). Over this range the uncertainties of the six intrinsic parameters shrink by factors of 13 to 69 (Methods), whereas the increase in signal-to-noise ratio alone would reduce them by a factor of only about 3.5. The additional gain arises from the slow evolution of the orbital frequencies as the compact object loses energy and angular momentum to gravitational radiation, an evolution that can be measured only by following the phase over the full year. The multipole structure of the black hole's spacetime, a surrounding dark-matter cloud or accretion disk, or a new fundamental field would all leave imprints on these orbital frequencies and their rates of change~\cite{ryan1995multipoles,eda2013dmspike,speri2023accretion,maselli2022fields}.

Finally, parallel-tempered Markov chain Monte Carlo integrates all surviving basins in the followed sector under the exact one-year likelihood. The resulting chain passes standard convergence tests, and explicit sampling of the luminosity distance then yields the 14-parameter posterior in Fig.~\ref{fig:posterior_geometry} (Methods). Because the basins are combined rather than reduced to the best one, the reported uncertainties include the spread between them.

\runinhead{Discussion}

We have solved the joint Bayesian inversion of a spinning EMRI described by the AAK waveform: starting from the complete prior and within the followed spin sector, all 14 parameters are recovered in a year-long, two-detector posterior. To our knowledge, no previous analysis has completed this inversion (Extended Data Table~\ref{tab:prior_work}). This matters because the EMRI science case for LISA and Taiji has so far rested on forecasts that assume the source parameters are known. Measuring the masses and spins of massive black holes, mapping how galactic nuclei capture compact objects, measuring the Hubble constant with dark sirens and, ultimately, testing strong-field gravity all require the source to be recovered from the data first. Within the AAK model, this forecast-level precision has now been achieved by a global analysis rather than assumed.

The approach divides the work in a simple way: neural networks decide where to look, and the exact likelihood decides what the data support. The search region acts as a truncation of the uniform prior, within which the exact detector likelihood determines the posterior, as in a standard Bayesian analysis~\cite{abbott2016properties}. Except for the spin direction and the correlated azimuthal phase, the posterior lies well inside the region's boundaries (Methods). Because the networks require only simulations, and the optimization and sampling steps only likelihood evaluations, the same framework can incorporate more accurate relativistic waveforms~\cite{katz2021few,wardell2023secondorder} and be applied to different detectors, such as TianQin~\cite{luo2016tianqin}. It should also carry over to other inference problems in which long observations yield high precision but the solution lies hidden in a large parameter space.

The present demonstration analyses one simulated source in stationary Gaussian noise. To keep the computation tractable, the exact-likelihood stage followed only one of the four spin sectors (Methods), so the measurement is conditional on that sector. The next steps are to propagate all candidate sectors and compare their evidence, to test the calibration of the credible intervals over many sources and noise realizations, and to include realistic noise, data gaps and overlapping signals. None of these steps changes the logic of the framework.

LISA and Taiji are planned to observe concurrently from around 2035, and LISA alone is expected to detect EMRIs at rates of a few to a few thousand per year~\cite{babak2017science}. Once extended to realistic data streams with overlapping sources, analyses of this kind could deliver, for each well-resolved event, the mass and spin of a massive black hole, the orbit on which a compact object reached it and its distance. Across many events, year-long phase tracking would thus become a census of massive black holes and of how galactic nuclei feed them across cosmic time, and an independent measurement of the expansion rate of the Universe.

Because the AAK describes a Kerr black hole in vacuum and no deviation from its Kerr quadrupole moment was sampled, the present posterior constrains the source under the Kerr hypothesis but cannot establish whether the spacetime is Kerr. Such a test requires deviation parameters, and where relevant environmental terms, to be sampled jointly with the 14 source parameters~\cite{barack2007offkerr}. Despite the additional degeneracies~\cite{zou2025bumpy}, this extension requires only simulations of the enlarged waveform model and no change to the neural-to-exact architecture, so year-long EMRI observations could then test whether massive black holes are the Kerr black holes of general relativity.

\FloatBarrier
\label{maintextend}
\bmhead{Data availability}
The data analysed in this study were simulated with the LISA Data Challenge (LDC)~\cite{baghi2022ldc} and Taiji Data Challenge (TDC)~\cite{ren2023tdc} simulation frameworks, using waveforms from \textsc{FastEMRIWaveforms}~\cite{katz2021few}. The injected parameters are listed in Extended Data Table~\ref{tab:final_intervals}. The Supplementary Information, which gives the network architectures and training, the simulation domain of every refinement round, the search and sampling settings and the intervals and marginal distributions of every continuation stage, is available at \url{https://github.com/aspirinliang/NEBULA-EMRI}.

\bmhead{Code availability}
The NEBULA-EMRI code, including the neural networks, the exact-likelihood search and sampling, and the scripts that reproduce the figures, will be made publicly available at \url{https://github.com/aspirinliang/NEBULA-EMRI}. Waveforms were generated with the publicly available \textsc{FastEMRIWaveforms} package~\cite{katz2021few}.

\bmhead{Acknowledgements}
We thank Alvin Chua for the EMRI workshop he organized at the National University of Singapore. This work is supported by the National Key R\&D Program of China under Grant No.~2025YFE0217300 and the International Partnership Program of the Chinese Academy of Sciences (Grant No.~025GJHZ2023106GC).

\bmhead{Competing interests}
The authors declare no competing interests.

\bibliography{references}

\clearpage
\section*{Methods}

\subsection*{Source model, prior and simulated observation}

We model the source with the augmented analytic kludge (AAK)~\cite{chua2017aak}, using the \texttt{Pn5AAKWaveform} implementation in \textsc{FastEMRIWaveforms} (FEW)~\cite{katz2021few}, which combines the AAK waveform construction with a fifth-order post-Newtonian inspiral trajectory. A generic spinning EMRI is described by 14 parameters: the masses $M$ and $\mu$ of the massive black hole and of the compact object; the dimensionless spin $a$; the initial semi-latus rectum $p_0$ (in units of $M$), eccentricity $e_0$ and inclination variable $Y_0\equiv\cos\iota_0=L_z/\sqrt{L_z^2+Q}$, where $L_z$ and $Q$ are the axial angular momentum and the Carter constant; the luminosity distance $D_L$; the ecliptic polar and azimuthal angles of the source, $(\theta_S,\phi_S)$, and of the spin, $(\theta_K,\phi_K)$, labelled $q_S$ and $q_K$ in FEW and in some figures; and the initial azimuthal, polar and radial orbital phases $\Phi_{\phi,0}$, $\Phi_{\theta,0}$ and $\Phi_{r,0}$. The masses are detector-frame (redshifted) masses. For a Planck cosmology the source lies at $z\approx0.40$, so its source-frame masses are smaller by $1+z\approx1.40$ and are known only as precisely as the redshift. The spin and the orbital elements are dimensionless and frame independent.

The prior, which we call the complete prior, is uniform and separable in these coordinates (Extended Data Table~\ref{tab:final_intervals}). The polar angles are uniform rather than isotropic, but across the 68\% intervals of $\theta_S$ and $\theta_K$ the isotropic weight $\sin\theta$ changes by only 0.6\% and 3.4\%, and a distance prior proportional to $D_L^2$ would move the distance posterior by less than 2\% of its standard deviation.

The observation was simulated with the LISA Data Challenge (LDC)~\cite{baghi2022ldc} and Taiji Data Challenge (TDC)~\cite{ren2023tdc} simulation frameworks. Waveforms are generated with a sampling interval $\Delta t=10\,\mathrm{s}$ over one sidereal year ($N_{\rm obs}=3{,}155{,}814$ samples) and projected onto the time-dependent responses of LISA ($2.5\times10^9\,\mathrm{m}$ arms on the ESA trailing orbit) and Taiji ($3.0\times10^9\,\mathrm{m}$ arms on its mission orbit) in the form of second-generation time-delay interferometry (TDI), including the orbital motion of the constellations and their time-varying arm lengths~\cite{prince2002tdi,tinto2014tdi,vallisneri2005synthetic,cornish2003response,amaro2017lisa,luo2020taiji,hu2017taiji,ruan2020network}. Inference uses the noise-orthogonal $A$ and $E$ channels of both detectors. Each response is multiplied by a Tukey window with taper fraction 0.05, zero-padded to $N_{\rm FFT}=4{,}194{,}304$ samples and Fourier transformed, and stationary Gaussian noise with the one-sided power spectral density (PSD) $S_{n,Ic}(f)$ of detector $I$ and channel $c$ is added independently in the four channels (Supplementary Section~1). The injected signal has optimal signal-to-noise ratios (SNRs) of 46.18 in LISA and 122.56 in Taiji, giving a network SNR of 130.98 (Extended Data Fig.~\ref{fig:simulated_emri_data}).

\subsection*{Network likelihood and distance profiling}

For two time series $x$ and $y$ in detector $I$ and TDI channel $c$, for example the data and a template, the noise-weighted inner product is
\begin{equation}
(x|y)_{Ic}
=\frac{4\Delta t}{N_{\rm FFT}}
\operatorname{Re}\sum_{k\in\mathcal{K}_{Ic}}
\frac{\widetilde{x}_{Ic,k}^{\,*}\,\widetilde{y}_{Ic,k}}{S_{n,Ic}(f_k)},
\label{eq:inner_product}
\end{equation}
where $\widetilde{x}_{Ic,k}$ is the unnormalized discrete Fourier transform of $x$ at frequency $f_k=k/(N_{\rm FFT}\Delta t)$, the asterisk denotes complex conjugation, $S_{n,Ic}$ is the one-sided noise PSD and $\mathcal{K}_{Ic}$ is the set of positive-frequency bins retained after masking the zeros of the TDI transfer functions. The continuous Fourier transforms $X$ and $Y$ of $x$ and $y$ satisfy $X(f_k)\approx\Delta t\,\widetilde{x}_{Ic,k}$ and $Y(f_k)\approx\Delta t\,\widetilde{y}_{Ic,k}$, and the bins are spaced by $\Delta f=1/(N_{\rm FFT}\Delta t)$. The prefactor $4\Delta t/N_{\rm FFT}=4(\Delta t)^2\Delta f$ therefore makes the sum approximate the standard continuous inner product $4\,\mathrm{Re}\int_0^\infty X^{*}(f)\,Y(f)/S_{n,Ic}(f)\,\mathrm{d}f$. Templates use the same taper and zero-padding as the data. The network inner product $(x|y)_{\mathrm{net}}$ sums equation~\eqref{eq:inner_product} over both detectors and both channels. At fixed values of the other 13 parameters $\boldsymbol{\theta}_{13}$, the waveform amplitude scales as $1/D_L$, so we write $h=\alpha\,h_0(\boldsymbol{\theta}_{13})$ with $\alpha=D_{\mathrm{ref}}/D_L$ and $h_0$ evaluated at $D_{\mathrm{ref}}=1\,\mathrm{Gpc}$. Up to a constant, the Gaussian log-likelihood is
\begin{equation}
\ln\mathcal{L}(\boldsymbol{\theta}_{13},D_L)
=-\tfrac12\bigl[d_d-2\alpha d_h+\alpha^2h_h\bigr],
\label{eq:coherent_likelihood}
\end{equation}
where $d$ denotes the data, $d_d=(d|d)_{\mathrm{net}}$ is the data power, $d_h=(d|h_0)_{\mathrm{net}}$ the overlap between the data and the reference template and $h_h=(h_0|h_0)_{\mathrm{net}}$ the power of the reference template, so that the bracket equals the noise-weighted squared residual $(d-h|d-h)_{\mathrm{net}}$. Maximizing over the distance within $[D_{\min},D_{\max}]$ gives the profile likelihood
\begin{equation}
\begin{aligned}
\ln\mathcal{L}_{\mathrm{prof}}(\boldsymbol{\theta}_{13})
&=-\tfrac12\bigl[d_d-2\widehat{\alpha}d_h+\widehat{\alpha}^{\,2}h_h\bigr],\\
\widehat{\alpha}
&=\operatorname{clip}\!\left(\frac{d_h}{h_h},\,\frac{D_{\mathrm{ref}}}{D_{\max}},\,\frac{D_{\mathrm{ref}}}{D_{\min}}\right),
\end{aligned}
\label{eq:profile_likelihood}
\end{equation}
where $\operatorname{clip}(x,a,b)=\min\{\max\{x,a\},b\}$ confines the best-fitting amplitude $d_h/h_h$ to the range allowed by the distance bounds $[D_{\min},D_{\max}]$ of the search domain. The differential-evolution (DE) searches and the 13-dimensional parallel-tempered Markov chain Monte Carlo (PTMCMC) runs use this profile likelihood. It keeps the exact $D_L^{-1}$ amplitude dependence while removing one sampled dimension. In the linear-signal approximation, maximizing and marginalizing over $D_L$ give the same covariance for the other parameters. Beyond this approximation, the two differ through non-Gaussian structure and through the distance bounds. The reported intervals therefore come from a separate run in which $D_L$ is sampled with equation~\eqref{eq:coherent_likelihood}.

\subsection*{Amortized full-prior network}

The first network, F0, is trained once over the complete prior and maps the observed data to a probability distribution over eight targets: the six intrinsic parameters $(M,\mu,a,p_0,e_0,Y_0)$, which fix the orbital frequencies, their secular evolution and the relative strength of the harmonics; $D_L$, which sets the amplitude; and $\log\rho_{\mathrm{net}}$, where $\rho_{\mathrm{net}}$ is the one-year network SNR. The sky position, the spin direction and the orbital phases act through weaker and more degenerate modulations. They keep their complete ranges at this stage and enter the later models once the intrinsic parameters have been constrained. The training set contains 50,000 sources drawn from the complete prior. Each time a waveform is loaded, it is rescaled to a network SNR drawn uniformly between 20 and 200, which fixes its distance, and a fresh noise realization is added, so that every source is seen at many signal strengths and in many noise realizations. An encoder of strided convolutions and ten MMDiT Transformer blocks, first trained as a heteroscedastic regressor, summarizes the data, and a conditional neural spline flow~\cite{durkan2019nsf} with 20 coupling transforms models the eight targets given this summary. Separating representation learning from density estimation follows related gravitational-wave analyses~\cite{chua2020neuralposteriors,dax2021realtime,dingobns2025nature}. Architectures and training schedules are given in Supplementary Section~2.

We tested F0 on 2,000 independent signals with network SNRs between 30 and 200, drawing 1,024 samples per signal. For each signal, the coordinate-wise central $1\sigma$, $2\sigma$ and $3\sigma$ intervals of the eight targets define a box, and the injected targets lie inside it for 16.2\%, 78.5\% and 98.0\% of the signals. For a calibrated Gaussian posterior, these fractions must lie between the values for eight independent targets, 4.7\%, 68.9\% and 97.9\%, which correlations can only raise~\cite{sidak1967rectangular}, and the single-target values, 68.3\%, 95.4\% and 99.7\%. The observed fractions lie within these ranges and give no indication of overconfident intervals. The median ratio of the complete target ranges to the $2\sigma$ box rises from 68.2 at SNR 30--50 to $3.43\times10^{9}$ at SNR 140--200 (Extended Data Fig.~\ref{fig:amortized_snr}). Because the box includes $\log\rho_{\mathrm{net}}$, this ratio measures contraction in the trained target space rather than a physical volume. On one NVIDIA H20 GPU, one forward pass of the encoder and 1,024 flow samples take less than 1\,s.

For the analysed observation, the central $2\sigma$ intervals of the seven physical targets define the first simulation domain, $M\in[9.62,9.90]\times10^{5}\,M_\odot$, $\mu\in[31.8,35.6]\,M_\odot$, $a\in[0.937,0.977]$, $p_0/M\in[11.94,12.18]$, $e_0\in[0.289,0.375]$, $Y_0\in[-0.394,-0.361]$ and $D_L\in[1.83,2.62]\,\mathrm{Gpc}$, which is $2.02\times10^{8}$ times smaller than the complete prior.

\subsection*{Sequential refinement and spin sectors}

Each later round simulates new sources inside the current domain, refits the conditional density to the observation and updates the domain. Every domain is an axis-aligned box $B_r=\prod_{j=1}^{14}[\ell^{(r)}_j,u^{(r)}_j]$, one interval per coordinate $j$, where $r$ labels the stage (F0, R1--R7, F14-S or F14-R). We measure it by its contraction relative to the complete prior $\prod_{j=1}^{14}[\ell^{\mathrm{prior}}_j,u^{\mathrm{prior}}_j]$ (Extended Data Table~\ref{tab:final_intervals}),
\begin{equation}
C_r=\prod_{j=1}^{14}\frac{u^{\mathrm{prior}}_j-\ell^{\mathrm{prior}}_j}{u^{(r)}_j-\ell^{(r)}_j}.
\label{eq:f0_r7_compression}
\end{equation}
Each factor is the ratio of the prior width of one coordinate to its retained width, so $C_r$ is the volume of the complete prior divided by the volume of $B_r$, and a contraction of $10^{8}$ means that the box occupies $10^{-8}$ of the prior volume. For the periodic angles, the retained width is the arc length kept. The quantity $C_r$ is a ratio of box volumes in this parameterization, not a posterior volume or an information measure. Over rounds R1--R7, the distance interval was narrowed first, the sky position and spin orientation were then restricted to one cell (R5), and the six intrinsic parameters were tightened last (R7), raising the contraction from $2.02\times10^{8}$ at F0 to $6.26\times10^{10}$. The orbital phases kept their complete ranges throughout (rules and bounds of each round in Supplementary Section~3).

High-resolution refinement then uses new networks trained on new simulations centred on a candidate found after R7 by particle-swarm optimization of a 90-day semicoherent statistic and coherent refinement. Under the one-year likelihood, three other branches that reached strong semicoherent maxima lie 3,700--5,000 log-likelihood units below the noise-only model, whereas the retained candidate lies 3,812 units above it, so the exact likelihood alone distinguishes them. Both high-resolution networks analyse, with a time--frequency Transformer encoder, the last 140.5 days of the year-long responses, which contain the highest-frequency and highest-amplitude part of the signal. The first, FlowA, learns the eleven non-phase parameters from 50,000 simulations in a domain centred on the candidate. The 5th and 95th percentiles of its samples define the simulation domain (F14-S) of the second network, Formal14, which learns all 14 parameters from 100,000 new simulations. With $\boldsymbol{\eta}_{11}$ the non-phase parameters, $\boldsymbol{\Phi}_0$ the phases and $\boldsymbol{z}$ the embedding of the data, Formal14 factorizes its density as
\begin{equation}
q_\psi(\boldsymbol{\eta}_{11},\boldsymbol{\Phi}_0\mid\boldsymbol{x})
=q_{\psi,\mathrm{np}}(\boldsymbol{\eta}_{11}\mid\boldsymbol{z})\,
q_{\psi,\mathrm{ph}}(\boldsymbol{\Phi}_0\mid\boldsymbol{\eta}_{11},\boldsymbol{z})\!,
\label{eq:refined_phase_factorization}
\end{equation}
where $\boldsymbol{x}$ denotes the whitened data analysed by the network and $\psi$ the network weights. The factor $q_{\psi,\mathrm{np}}$ is the density of the non-phase parameters given the embedding, modelled by a neural spline flow, and $q_{\psi,\mathrm{ph}}$ is the density of the three phases given the non-phase parameters and the embedding, modelled by a conditional circular spline flow on the three-torus. This factorization does not assume that the phases are independent of the other parameters. Formal14 returns 4,096 proposals for the observation.

For each non-phase coordinate, the search bounds are the 0.5\% and 99.5\% quantiles of these proposals, intersected with the F14-S interval. The 3,968 proposals whose spin directions lie inside F14-S form a ring (Fig.~\ref{fig:neural_topology}b), which $k$-means clustering of the spin unit vectors divides into four sectors of 893, 1,056, 825 and 1,194 members. Differential evolution run within each of the four sectors reaches its highest likelihood in the 825-member sector, which contains the injected spin direction. We follow this sector and set the $(\theta_K,\phi_K)$ bounds to its central 99\% intervals. The resulting search domain, F14-R, is $5.41\times10^{17}$ times smaller than the complete prior, and F14-S is $1.39\times10^{16}$ times smaller (Fig.~\ref{fig:neural_topology}a and Extended Data Table~\ref{tab:final_intervals}). The distance is profiled over the FlowA interval. From the 697 proposals of the followed sector that satisfy all eleven non-phase bounds, 22 representatives are selected in the ten coordinates other than distance and phase, and each is paired with 128 scrambled Sobol points over the phase torus, giving 2,816 initial states. The neural density selects these starting points but never enters the likelihood.

Where candidate regions were compared by their likelihood, as for the particle-swarm branches after R7 and the four spin sectors, the likelihood alone selected the region to follow.

\subsection*{Exact-likelihood search and two-detector screening}

A one-year search started directly from the region identified by the networks does not reach the posterior. Three blocks of DE under the one-year profile likelihood, which vary first the three phases, then the ten non-phase coordinates and finally all 13 coordinates (Supplementary Section~4), end on secondary maxima: their best candidates lie 9,560.1, 9,341.0 and 9,250.0 log-likelihood units below the injected parameters, although every coordinate is close to its injected value. Relative to the noise-only model, the injected parameters lie 8,657 units above and the best candidate 593.5 units below. The analysis therefore restarts from 30 days of data.

The terminal 13-dimensional populations (R13 in Fig.~\ref{fig:exact_handoffs}) initialize independent 30-day DE searches in each detector under a phase-relaxed statistic, which maximizes the likelihood of detector $I$ over a common amplitude, within the distance bounds, and over one additional phase per data segment, and compares the result with noise alone:
\begin{equation}
\begin{aligned}
\mathcal S_I(\boldsymbol{\theta})=2\Bigl[&
\max_{\alpha,\{\delta\phi_s\}}\ln\mathcal L_I\bigl(d_I\mid\alpha h_I(\boldsymbol{\theta};\{\delta\phi_s\})\bigr)\\
&-\ln\mathcal L_I(d_I\mid h_I=0)\Bigr].
\end{aligned}
\label{eq:mode_score_interpretation}
\end{equation}
Here $d_I$ denotes the data of detector $I$, $\mathcal{L}_I$ the likelihood of equation~\eqref{eq:coherent_likelihood} restricted to that detector, $h_I(\boldsymbol{\theta};\{\delta\phi_s\})$ its template at the reference distance $D_{\mathrm{ref}}$ for the 13 parameters $\boldsymbol{\theta}$ other than the distance, with an extra phase shift $\delta\phi_s$ applied in data segment $s$, and $\alpha=D_{\mathrm{ref}}/D_L$ the amplitude scale, confined to the distance bounds. The second term is the log-likelihood of noise alone, so a positive score means that the template fits that detector's data better than noise alone. Because the distance bounds keep the amplitude away from zero, a poor template can score negatively. We use a single 30-day segment per detector, and the additional phase is common to both channels and distinct from the three orbital phases. Each detector search returns 2,048 candidates, and all 4,096 are evaluated in both detectors (Extended Data Fig.~\ref{fig:mode_detector_consistency}). Of these, 622 have positive scores in both detectors, 417 only in LISA, 162 only in Taiji and 2,895 in neither. The 622 are ranked by the smaller of their two scores, each normalized by its maximum over the candidate bank, so that a high score in one detector cannot compensate for a low score in the other. For reference, the injected parameters score $(59.22,257.08)$ in LISA and Taiji, or $(1.016,1.266)$ after normalization. $k$-means clustering divides the 622 into 16 groups, whose best-ranked members anchor 16 populations of 128 particles for 20 generations of joint 30-day DE with one waveform and one distance for both detectors. The screening thus shapes the initial populations of the joint search but changes neither its bounds nor its target. We have not measured its benefit against an unscreened initialization of equal cost.

\subsection*{Coherence continuation and one-year sampling}

The joint search is continued over coherent observations of 30, 60, 180 and 360 days and then one year. For duration $T$, the target is
\begin{equation}
\begin{aligned}
\ell_T^{\mathrm{prof}}(\boldsymbol{\theta}_{13})
&=\max_{D_{\min}\leq D_L\leq D_{\max}}\ln\mathcal L_T(d_T\mid\boldsymbol{\theta}_{13},D_L)\!,\\
\widetilde p_T(\boldsymbol{\theta}_{13}\mid d_T)
&\propto\pi_{13}(\boldsymbol{\theta}_{13})\exp\!\bigl[\ell_T^{\mathrm{prof}}(\boldsymbol{\theta}_{13})\bigr],
\end{aligned}
\label{eq:continuation_target}
\end{equation}
where $d_T$ denotes the data restricted to a coherent duration $T$, $\mathcal{L}_T$ the two-detector likelihood of equation~\eqref{eq:coherent_likelihood} evaluated on these data, $\ell_T^{\mathrm{prof}}$ the maximum of $\ln\mathcal{L}_T$ over the distance, $\pi_{13}$ the uniform prior over the fixed search domain and $\widetilde p_T$ the resulting profile posterior at duration $T$. Detector responses and Fourier transforms are rebuilt for each $T$, and the two detectors share one waveform and one distance. At each finite duration, DE initializes a short PTMCMC run, whose likelihood-ranked and mutually separated cold-chain representatives seed the next duration. Likelihood values from different durations are not compared, because the data differ. At one year, three PTMCMC runs with eight, twelve and sixteen temperatures follow, each seeded from the cold chain of the previous one. The final run uses 256 walkers per temperature and 2,000 iterations, and discarding the first 1,000 leaves 256,000 correlated cold-chain states (settings of all runs in Supplementary Section~5). All runs use the ensemble stretch move with an adaptive temperature ladder, and only the cold chain enters the reported distributions.

In the final run, the cold-chain stretch acceptance is 20.4\%, and swaps between adjacent temperatures are accepted at rates between 27.2\% and 92.7\% (Extended Data Fig.~\ref{fig:ptmcmc_diagnostics}). After burn-in, the cold-chain likelihood is stationary, all 16 seeded basins remain occupied, the basin occupancies of the two halves of the retained chain differ by a total-variation distance of 4.4\%, and the largest split-$\widehat{R}$ over the 13 coordinates is 1.0085. The basins differ mainly in the spin direction and the azimuthal phase, which span 52--91\% of their search widths, whereas the intrinsic marginals of the combined posterior are unimodal (Fig.~\ref{fig:posterior_geometry}).

Extending the coherent duration sharpens the intrinsic parameters far beyond the gain in SNR. At 30 days, the 68\% half-widths of $(M/10^{6}M_\odot,\,\mu/M_\odot,\,a,\,p_0/M,\,e_0,\,Y_0)$ are $(5.9\times10^{-4},\,0.067,\,7.2\times10^{-4},\,3.4\times10^{-3},\,8.5\times10^{-4},\,8.0\times10^{-4})$. The one-year explicit-distance intervals are 13--69 times narrower, whereas the SNR alone grows by $\sqrt{365/30}\approx3.5$. The one-year profile sample, which is truncated as described below, gives factors of 23--120. The additional gain comes from the secular evolution of the orbital frequencies, which can be resolved only by following the phase over the full year. The spin-direction widths do not shrink monotonically, because they approach the limits of the sector's search domain. Non-periodic marginals are summarized by their median and central 68.27\% interval, and the phases by offsets from the circular means of the final profile chain, $(\Phi_{\phi,0},\Phi_{\theta,0},\Phi_{r,0})_{\rm ref}=(3.780632,0.852990,6.119448)\,\mathrm{rad}$. The intervals and marginal distributions of every stage are given in Supplementary Section~5.

\subsection*{Explicit-distance sampling and reported intervals}

Profiling optimizes over the distance instead of integrating over it, and the profile chain is confined to the search domain. To obtain the reported posterior, we therefore sample $D_L$ explicitly with the likelihood of equation~\eqref{eq:coherent_likelihood}, keeping the data, waveform, noise weighting and prior form unchanged. In a first explicit-distance run, samples accumulated at the domain boundaries in $\mu$, $D_L$ and $\phi_K$, so we widened all eleven non-phase intervals by a factor of 1.5 about their centres ($D_L\in[2.18133,2.70303]\,\mathrm{Gpc}$; phases unchanged) and restarted from the final profile chain. The run uses 16 temperatures and 256 walkers per temperature. It was stopped after 914 of the 2,000 planned iterations, and we use iterations 451--900, which give 115,200 cold-chain states (Fig.~\ref{fig:posterior_geometry} and Extended Data Table~\ref{tab:final_intervals}).

The explicit-distance intervals of the six intrinsic parameters are 1.73--1.78 times wider than those of the profile chain, whereas those of the sky angles agree to within 2\%. Profiling cannot explain the difference, because it gives the same covariance as marginalization in the linear-signal approximation. The cause is the boundary of the original domain. Its lower $\mu$ bound, $33.1108\,M_\odot$, lies only $0.0018\,M_\odot$ below the injected value but above the explicit-distance median of $\mu$, $33.1104\,M_\odot$, and roughly 40\% of the explicit-distance samples lie below the original distance limit of 2.268\,Gpc, where the profiled amplitude is clipped. Because the six intrinsic parameters are tightly correlated (Fig.~\ref{fig:posterior_geometry}), the $\mu$ bound truncates the profile chain along the principal degeneracy, which explains both the asymmetry of the profile intervals and their common offset of about 0.7 standard deviations from the explicit-distance medians. The profile chain therefore serves to locate and connect the basins, and all reported intervals, including the precisions in Table~\ref{tab:physics_reach}, come from the explicit-distance run. These precisions are the central 68\% half-widths of the explicit-distance marginals divided by their medians, or absolute half-widths for the sky angles.

Because that run was stopped early, its convergence is less firmly established than that of the profile chain. Three observations support the reported intervals. The marginals are smooth and unimodal except for the spin direction and $\Phi_{\phi,0}$ (Fig.~\ref{fig:posterior_geometry}). Every median lies within 0.75 standard deviations of the corresponding profile median. Apart from $\theta_K$ and $\phi_K$, no coordinate has more than 0.001\% of its samples within 2\% of the widened boundaries. The spin azimuth and $\Phi_{\phi,0}$, which are strongly correlated through the ring degeneracy, remain bounded by the domain of the followed sector, so their intervals are conditional on that sector, as is the measurement as a whole.

All six intrinsic parameters and $D_L$ lie inside their 68\% intervals. The sky angles $\theta_S$ and $\phi_S$ are offset by 2.1 and 2.4 standard deviations, and $\phi_K$ and $\Phi_{\phi,0}$ lie just outside their intervals, at 1.2 and 1.3 half-widths from the median. The same four coordinates lie outside their intervals in the profile chain. Four of fourteen coordinates outside their 68\% intervals is in line with the 4.4 expected for calibrated intervals, and with fourteen coordinates an offset of at least 2.4 standard deviations in one of them occurs in about 20\% of noise realizations. A single realization cannot test calibration, which requires repeated injections.

\subsection*{Comparison with previous EMRI analyses}

Extended Data Table~\ref{tab:prior_work} compares NEBULA-EMRI with previous end-to-end EMRI analyses that deliver posterior uncertainties. Strub et al.~\cite{strub2025search} recovered a non-spinning eccentric EMRI from wide mass and eccentricity priors in 0.44~yr of LISA data, with a time--frequency search followed by Markov chain Monte Carlo (MCMC) over 11 parameters started at the search maximum. Cole et al.~\cite{cole2026sequential} used truncated marginal neural ratio estimation to contract the 11-dimensional prior of non-spinning EMRIs by a factor of $10^{6}$--$10^{7}$ in 0.1~yr of data, obtaining one-dimensional marginal proposal distributions rather than a joint posterior. Our previous flow-matching MCMC study~\cite{liang2025fmmcmc} recovered the joint posterior of four intrinsic parameters of a spinning AAK source from broad priors in 60 days of Taiji data, with the extrinsic parameters initialized near their injected values, and showed that PTMCMC started from broad priors is trapped by secondary maxima. Each of these analyses addresses a simplified version of the problem: it omits the spin, restricts the parameters inferred from broad priors or analyses at most half a year of single-detector data.

Searches that return point estimates have also advanced~\cite{gair2008semicoherent,babak2009algorithm,zou2026endtoend}. Most recently, Zou et al.~\cite{zou2026endtoend} coherently searched all 14 parameters of half-year, single-detector signals generated with the original analytic kludge, whose orbital frequencies, unlike those of the AAK, are not mapped onto Kerr geodesics~\cite{chua2017aak}. They reported fractional errors of $10^{-3}$--$10^{-2}$ on the phase-evolution parameters, about 3\% on the distance and below 0.1\,rad on the sky position, without recovering the spin orientation or the initial phases. A point estimate carries neither credible intervals nor parameter covariances, so such results are not included in Extended Data Table~\ref{tab:prior_work}. Nor does it guarantee that a posterior can be computed: in our analysis, optimization candidates close to the injection in every coordinate still lay 9,250--9,560 log-likelihood units below it under the one-year likelihood, so a sampler started from them would explore a secondary basin~\cite{liang2025fmmcmc}. Reaching the posterior required two-detector screening, coherence continuation and multibasin sampling. Numerical comparisons between studies should account for their different SNRs, observation times, waveform models and detector configurations.

\FloatBarrier
\clearpage
\onecolumn
\section*{Extended Data}
\setcounter{figure}{0}
\setcounter{table}{0}
\renewcommand{\figurename}{Extended Data Fig.}
\renewcommand{\tablename}{Extended Data Table}
\renewcommand{\theHfigure}{ED\arabic{figure}}
\renewcommand{\theHtable}{ED\arabic{table}}

\begin{table}[!htbp]
\centering
\caption{\textbf{End-to-end EMRI analyses that deliver posterior uncertainties: the inference problem solved and the physics accessible.} The upper block describes the inference problem, and the lower block lists the physical measurements that each posterior supports. Parameters are those inferred from broad priors. $\checkmark$, accessible; $\times$, not accessible.}
\label{tab:prior_work}
\small
\begin{tabularx}{\textwidth}{>{\raggedright\arraybackslash}p{0.28\textwidth} >{\centering\arraybackslash}X >{\centering\arraybackslash}X >{\centering\arraybackslash}X >{\centering\arraybackslash}X}
\toprule
 & Strub et al.~\cite{strub2025search} & Cole et al.~\cite{cole2026sequential} & Liang et al.~\cite{liang2025fmmcmc} & \textbf{This work}\\
\midrule
\multicolumn{5}{l}{\textit{Inference problem}}\\
Generic spinning source & $\times$ & $\times$ & $\checkmark$ & $\checkmark$\\
Parameters inferred from broad priors & 11 & 11 & 4 & \textbf{14}\\
Prior & Wide mass and eccentricity priors & Wide 11-dimensional prior & Broad intrinsic; extrinsic near injection & \textbf{Complete prior (Extended Data Table~\ref{tab:final_intervals})}\\
Coherent data & 0.44\,yr & 0.1\,yr & 60\,d & \textbf{1\,yr}\\
Detectors & LISA & LISA & Taiji & \textbf{LISA + Taiji}\\
Waveform & Relativistic, non-spinning & Relativistic, non-spinning & AAK & \textbf{AAK}\\
Posterior & Joint, 11 parameters & One-dimensional marginal proposals & Joint, 4 intrinsic parameters & \textbf{Joint and multibasin, all 14 parameters}\\
\midrule
\multicolumn{5}{l}{\textit{Physics accessible from the posterior}}\\
Black-hole spin & $\times$ & $\times$ & $\checkmark$ & $\checkmark$\\
Capture channel (eccentricity and inclination) & $\times$ & $\times$ & $\times$ & $\checkmark$\\
Dark siren (distance and sky position) & $\checkmark$ & Marginals only & $\times$ & $\checkmark$\\
Joint covariance of all source parameters & Non-spinning only & $\times$ & $\times$ & $\checkmark$\\
Confirmation in an independent detector & $\times$ & $\times$ & $\times$ & $\checkmark$\\
\bottomrule
\end{tabularx}
\end{table}

\begin{figure}[!htbp]
  \centering
  \includegraphics[width=0.55\textwidth]{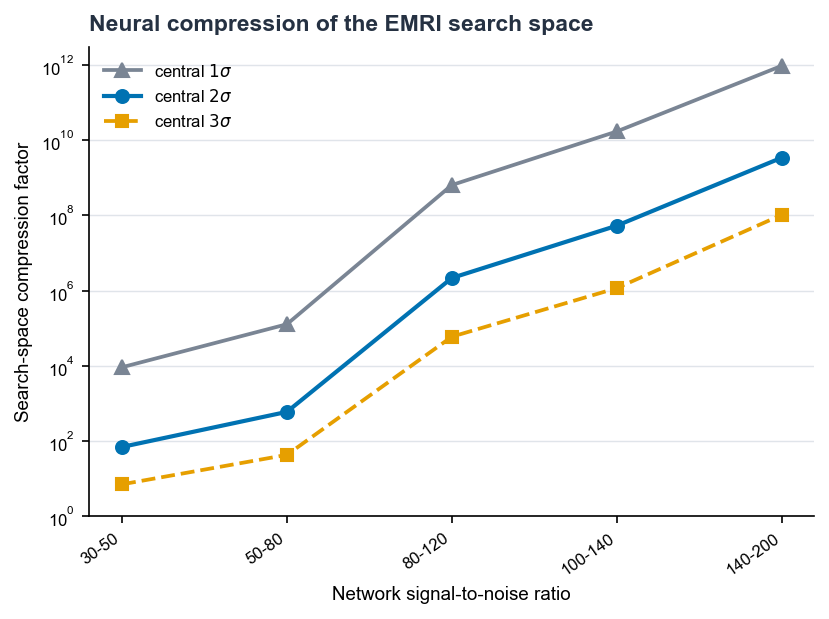}
  \caption{\textbf{Contraction of the amortized search region grows with signal-to-noise ratio.} Median contraction of the search support produced by the amortized network for 2,000 independent simulated signals with different source parameters and fresh noise, grouped in windows of network signal-to-noise ratio (SNR). The 80--120 and 100--140 windows overlap. For each signal, 1,024 draws from the conditional flow define a box from the coordinate-wise central $1\sigma$, $2\sigma$ or $3\sigma$ intervals of the eight targets (the seven physical parameters and the logarithm of the network SNR), and the contraction is the ratio of the complete target ranges to this box. Across all signals, the injected targets lie inside the $1\sigma$, $2\sigma$ and $3\sigma$ boxes in 16.2\%, 78.5\% and 98.0\% of cases. Lines are guides to the eye.}
  \label{fig:amortized_snr}
\end{figure}

\begin{figure}[!htbp]
\centering
\includegraphics[width=0.92\textwidth]{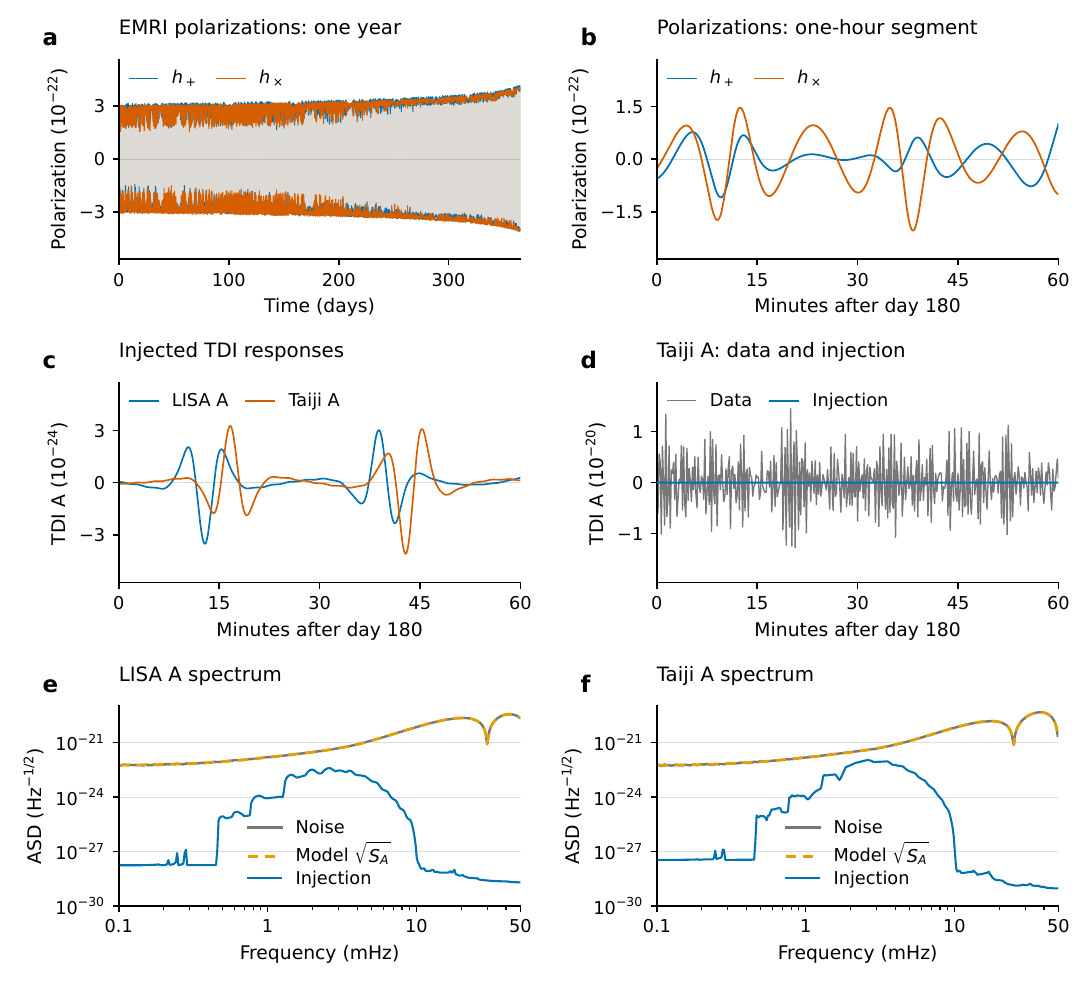}
\caption{\textbf{Simulated EMRI signal and detector noise.}
\textbf{a}, Year-long envelopes of the incident polarizations $h_+$ and $h_\times$ at the injected parameters. Bands show the minimum and maximum in each of 2,000 time blocks.
\textbf{b}, Both polarizations during one hour beginning at day 180.
\textbf{c}, Noiseless LISA and Taiji $A$-channel TDI responses in the same interval.
\textbf{d}, Taiji $A$-channel data and injected response on the same amplitude scale.
\textbf{e,f}, LISA and Taiji $A$-channel amplitude spectra of the injected response and of the noise realization, compared with the noise model. Curves show square roots of powers averaged in 320 logarithmic frequency bins, with the model PSD averaged over the same bins. For interior bins the periodogram is $P_x(f_k)=2\Delta t\,|\widetilde{x}_k|^2/N_{\rm FFT}$.
The $A$ channels are shown as examples. Inference uses both $A$ and $E$ in each detector. The unwhitened time series is dominated by noise, whereas coherent analysis accumulates the signal over the full year.}
\label{fig:simulated_emri_data}
\end{figure}

\begin{table}[!htbp]
\centering
\caption{\textbf{Prior, search domain and posterior intervals.} Prior: the complete prior, uniform in each parameter. Search domain: the region F14-R used by the exact-likelihood search and the profile chain. Injected: the simulation input. Profile: the one-year, sixteen-temperature profile-likelihood chain (256,000 states). Explicit distance: the run with $D_L$ sampled in a domain widened by a factor of 1.5 about the same centres (115,200 states; Fig.~\ref{fig:posterior_geometry}). Entries give the median and the central 68.27\% interval. Precision: half-width of the explicit-distance interval divided by the median for $M$--$D_L$, and absolute half-width (rad) for the angles. Masses are in the units of the first column and angles in radians. Phases are offsets from $(3.780632,0.852990,6.119448)$\,rad, whereas their prior and search domain refer to the phases themselves. The lower $\mu$ bound and the distance limit truncate the profile chain (Methods).}
\label{tab:final_intervals}
\footnotesize
\setlength{\tabcolsep}{3pt}
\renewcommand{\arraystretch}{1.45}
\begin{tabular}{@{}l c c c c c c@{}}
\toprule
Parameter & Prior & Search domain & Injected & Profile (13D) & Explicit distance (14D) & Precision \\
\midrule
$M/(10^6M_\odot)$ & $[0.5,1.0]$ & $[0.9767833,0.9794250]$ & $0.97742962$ & $0.9774322^{+0.0000079}_{-0.0000140}$ & $0.9774451^{+0.0000188}_{-0.0000192}$ & $1.9\times10^{-5}$ \\
$\mu/M_\odot$ & $[10,50]$ & $[33.11076,33.37859]$ & $33.112520$ & $33.1126^{+0.0023}_{-0.0014}$ & $33.1104^{+0.0033}_{-0.0032}$ & $9.7\times10^{-5}$ \\
$a$ & $[0.1,0.999]$ & $[0.941609,0.953381]$ & $0.94544326$ & $0.9454428^{+0.0000160}_{-0.0000098}$ & $0.9454266^{+0.0000234}_{-0.0000226}$ & $2.4\times10^{-5}$ \\
$p_0/M$ & $[10,16]$ & $[12.002577,12.028257]$ & $12.0177123$ & $12.01771^{+0.00019}_{-0.00011}$ & $12.01753^{+0.00027}_{-0.00026}$ & $2.2\times10^{-5}$ \\
$e_0$ & $[0.1,0.9]$ & $[0.314074,0.324895]$ & $0.32030659$ & $0.3203026^{+0.0000056}_{-0.0000086}$ & $0.3203116^{+0.0000122}_{-0.0000125}$ & $3.9\times10^{-5}$ \\
$Y_0$ & $[-1,1]$ & $[-0.385872,-0.372277]$ & $-0.38479048$ & $-0.384792^{+0.000032}_{-0.000019}$ & $-0.384823^{+0.000046}_{-0.000044}$ & $1.2\times10^{-4}$ \\
$D_L/\mathrm{Gpc}$ & $[0.45,3.0]$ & $[2.26828,2.61608]$\textsuperscript{a} & $2.281054$ & profiled & $2.2720^{+0.0179}_{-0.0184}$ & $8.0\times10^{-3}$ \\
$\theta_S$ & $[0,\pi]$ & $[2.04141,2.36553]$ & $2.2516818$ & $2.2442^{+0.0036}_{-0.0037}$ & $2.2439^{+0.0038}_{-0.0036}$ & $3.7\times10^{-3}$ \\
$\phi_S$ & $[0,2\pi)$ & $[0.68328,0.93374]$ & $0.7913202$ & $0.7984^{+0.0030}_{-0.0029}$ & $0.7986^{+0.0030}_{-0.0030}$ & $3.0\times10^{-3}$ \\
$\theta_K$ & $[0,\pi]$ & $[1.15476,1.54627]$ & $1.4189733$ & $1.451^{+0.040}_{-0.098}$ & $1.431^{+0.058}_{-0.129}$ & $0.093$ \\
$\phi_K$ & $[0,2\pi)$ & $[3.39967,4.31087]$ & $4.2494406$ & $3.79^{+0.36}_{-0.29}$ & $3.78^{+0.42}_{-0.34}$ & $0.38$ \\
$\Delta\Phi_{\phi,0}$ & $[0,2\pi)$ & $[0,2\pi)$ & $0.641799$ & $-0.03^{+0.50}_{-0.41}$ & $-0.05^{+0.59}_{-0.51}$ & $0.55$ \\
$\Delta\Phi_{\theta,0}$ & $[0,2\pi)$ & $[0,2\pi)$ & $0.010602$ & $0.000^{+0.029}_{-0.030}$ & $-0.014^{+0.034}_{-0.033}$ & $0.034$ \\
$\Delta\Phi_{r,0}$ & $[0,2\pi)$ & $[0,2\pi)$ & $0.015264$ & $0.000^{+0.025}_{-0.025}$ & $0.017^{+0.030}_{-0.030}$ & $0.030$ \\
\bottomrule
\end{tabular}
\par\smallskip
\raggedright\footnotesize
\textsuperscript{a}Profiling interval of the 13-parameter runs. The explicit-distance run samples $D_L\in[2.18133,2.70303]$\,Gpc.
\end{table}

\begin{figure}[!htbp]
\centering
\includegraphics[width=0.85\textwidth]{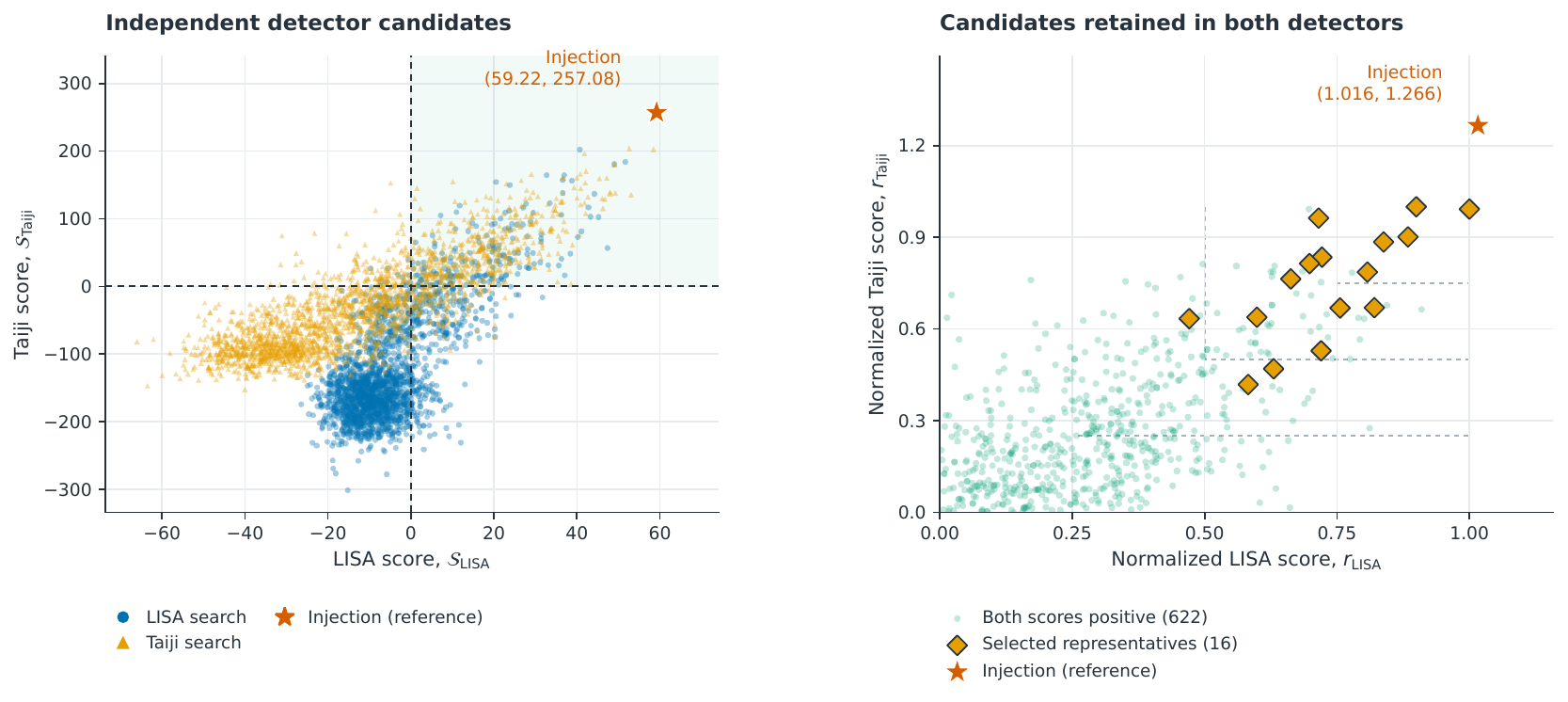}
\caption{\textbf{Two-detector screening of the 30-day candidates.} Left, LISA and Taiji scores (equation~\eqref{eq:mode_score_interpretation}) of the 4,096 candidates from the independent searches. Blue circles and gold triangles mark candidates found in the LISA and Taiji searches. A score is twice the log-likelihood improvement over noise alone after maximizing one additional phase and a bounded amplitude. Dashed lines at zero enclose the 622 candidates that are positive in both detectors. Right, the retained candidates after normalization by the bank maximum in each detector. Gold diamonds mark the 16 representatives, and L-shaped contours show $\min(r_{\mathrm{LISA}},r_{\mathrm{Taiji}})=0.25$, 0.50 and 0.75. Vermilion stars mark the injected parameters, with scores $(59.22,257.08)$ and normalized scores $(1.016,1.266)$. The injection is not part of the bank used for normalization, so its normalized scores can exceed unity.}
\label{fig:mode_detector_consistency}
\end{figure}

\begin{figure}[!htbp]
\centering
\includegraphics[width=0.78\textwidth]{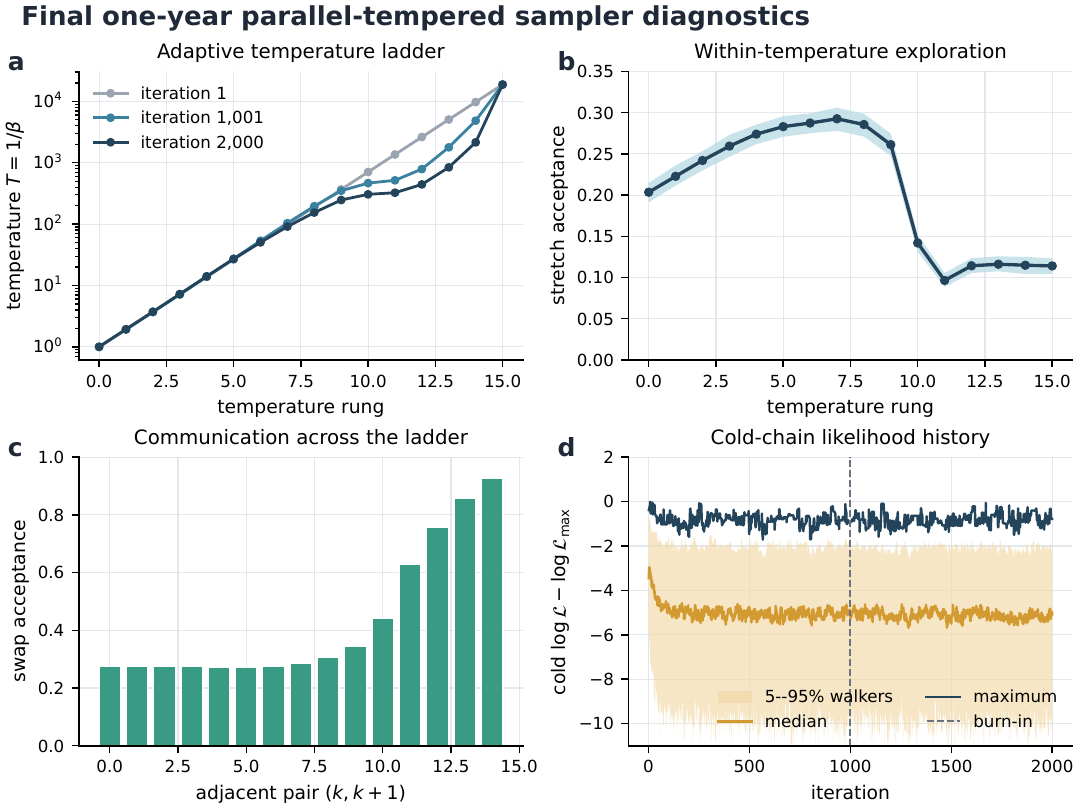}
\caption{\textbf{Diagnostics of the final one-year profile-likelihood PTMCMC run.} \textbf{a}, Adaptive temperature ladder at the first, 1,001st and last iteration. \textbf{b}, Stretch-move acceptance at each temperature. \textbf{c}, Swap acceptance between adjacent temperatures. \textbf{d}, Cold-chain log likelihood relative to its maximum: maximum and median over walkers and the 5--95\% range. The dashed line marks the end of the 1,000 discarded iterations.}
\label{fig:ptmcmc_diagnostics}
\end{figure}

\end{document}